\documentstyle[12pt]{report}
\def\lsim{\lower0.6ex\vbox{\hbox{$ \buildrel{\textstyle <}\over{\sim}\ $}}}
\def\rsim{\lower0.6ex\vbox{\hbox{$ \buildrel{\textstyle >}\over{\sim}\ $}}}
\begin{document}
\noindent
         \hfill OSU-TA-4/95\\
 \vspace{ 1.3cm}
\begin{center}
\begin{Large}
\begin{bf}
COSMIC LITHIUM:  GOING UP \\
OR COMING DOWN?\\
\end{bf} \end{Large}
\end{center}
\begin{center}
\begin{large}
GARY STEIGMAN\\
\end{large}
 \vspace{0.3cm}
Departments of Physics and Astronomy\\
The Ohio State University\\
174 West 18th Avenue, Columbus, OH 43210, USA\\
\end{center}
 \vspace{2.5cm}
\vfil\eject
 \pagenumbering{roman}
\global\advance\count0 by -3
\centerline{\bf ABSTRACT}
\begin{quotation}
\noindent
Observations of interstellar lithium provide a valuable complement to
studies of
lithium in Pop I and Pop II stars.   Large corrections for unseen LiII and
for non-gas
phase lithium have provided obstacles to using interstellar data for abundance
determinations.  An approach to surmounting these difficulties is proposed
and is
applied to the Galaxy and the LMC.  The key is that since potassium and
lithium behave
similarly   regarding  ionization and depletion, their observed ratio
(LiI/KI) can be used to
probe the abundance and evolution of lithium.  For ten lines-of-sight in the
interstellar medium of the Galaxy (ISM) the Li/K ratio observed $(\log
(N_{Li}/N_K)_{ISM} = - 1.88 \pm 0.09)$ is entirely consistent with the
solar system
value $(\log (N_{Li}/N_{K})_{\odot} = - 1.82 \pm 0.05)$.  The absence of
LiI in front of
SN87A in the LMC, coupled with the observed KI, corresponds to an upper
bound (at $\rsim 95 \%
\ CL$) of $\log (N_{Li}/N_K)_{LMC} < - 0.3 +   \log (N_{Li}/N_K)_{ISM}.$
This low upper
bound to LMC lithium suggests that cosmic lithium is on its way up from a
primordial abundance
lower, by at least a factor of two, than the present Pop I value of
$[Li]_{PopI} \equiv 12 + \log
(Li/H)_{PopI} = 3.2 \pm 0.1$.

\end{quotation}
\vfil\eject
\global\advance\count0 by -2
 \pagenumbering{arabic}

\noindent
 INTRODUCTION

Cosmic lithium provides a valuable probe of stellar structure and evolution, of
Galactic chemical evolution and, of cosmology. In particular its primordial
abundance
is of special importance for testing Big Bang Nucleosynthesis (BBN)
providing, as it
does, a tool for discriminating between homogeneous (``standard") BBN
(e.g., Boesgaard
\& Steigman 1985; Walker et al.\ 1991; Smith, Kawano \& Malaney 1993) and
inhomogeneous
BBN (e.g., Alcock, Fuller \& Mathews 1987; Applegate, Hogan \& Scherrer
1988; Malaney \&
Fowler 1988; Terasawa \& Sato 1990; Kurki-Suonio \& Matzner 1990; Thomas et
al.\ 1994).
There are two observational approaches to primordial lithium: via stars and
via the
interstellar medium (ISM).  Each has its assets and its liabilities. In
recent years
the ``traditional" approach to primordial lithium has been to utilize stellar
observations, especially the metal-poor $([Fe/H] ~\lsim - 1.3)$, warm $(T
{}~\rsim 5700K)$
Pop II stars of the ``Spite Plateau" (Spite \& Spite 1982).  The vast
majority of such
stars have a lithium abundance $[Li]_{PopII} \equiv 12 + log (Li/H)_{PopII}
= 2.1 \pm
0.1$ which is independent of metallicity for $- 3.5 ~\lsim ~[Fe/H] ~\lsim
-1.3$ (Spite,
Maillard \& Spite 1984; Spite \& Spite 1986; Rebolo, Molaro \& Beckman
1988; Hobbs \&
Pilachowski 1988; Thorburn 1994).  This metallicity plateau is evidence for
the Pop II
lithium abundance having the primordial value.

However, there are some complications.  It has been long known that Pop I stars
 \break
\hbox{$([Fe/H]_{PopI}~ \approx 0)$} of similar temperatures to those in the
Spite Plateau
have depleted their surface lithium, often by a very large factor.  For
example, although
the solar system (meteoritic) abundance of lithium is $[Li]_{\odot} = 3.31
\pm 0.04$
(Grevesse \& Anders 1989), in the solar photosphere $[Li] = 1.2 \pm 0.1$.
Lithium is
easily destroyed in stars, burning at the low temperature of $\sim 2 \times
10^6 $ K. \ It
will only survive on the surface of those stars whose convective layers are
sufficiently thin that the surface material is not exposed to such
temperatures.  The
solar surface depletion is a general trend seen in the cooler Pop I stars
in open
clusters and the field (see, e.g., the discussion in Boesgaard \& Steigman
1985 and
references therein).  The surprise of the Spite Plateau is that the lithium
appears to
survive in stars which have had much longer to burn it away.  Could it be
that the
primordial abundance of lithium was much larger (larger, even, than the Pop
I/solar
values) and, the structure and evolution of the Pop II stars has conspired
to destroy lithium
to the level of  the Spite Plateau (Mathews et al. 1990)?  Standard
(i.e., non-rotating)
stellar  models for Pop II stars in the Spite Plateau in fact predict very
little lithium
depletion and Chaboyer et al. (1992) conclude on the basis of such models
that the
primordial abundance is $[Li]_P = 2.15$.  It should be noted that the very
recent,
extensive data set and analysis of Thorburn (1994) suggests Pop II lithium
abundances which are
systematically higher than earlier results by $\sim 0.2$ dex.  For
diffusive models,
Chaboyer et al. (1992) conclude that modest depletion $( \lsim 0.1$ dex)
may have
occurred.  In contrast, for rotating models Pinsonneault, Deliyannis \&
Demarque
(1992) find that large depletion $( \sim 0.7 - 1.0$ dex) (see also,
Charbonell \&
Vauclair 1992) is predicted.  Although observations of \ $^6$Li in a few
Pop II stars
(Smith, Lambert \& Nissen 1992; Hobbs \& Thorburn 1994) would appear to
argue against
such large lithium destruction, the situation at present is unclear (M.
Pinsonneault,
Private Communication).

Another complication on the path to primordial lithium is early production
in cosmic
ray nucleosynthesis (Steigman \& Walker 1992).  Observations of $^6$Li, Be
and B in
Pop II stars (Ryan et al.\ 1990; Gilmore, Edvardsson \& Nissen 1992; Ryan
et al.\ 1992;
Duncan, Lambert \& Lemke 1992) provide evidence for cosmic ray nucleosynthesis
(Steigman et al.\ 1993) but, there are more free/adjustable parameters than
data points,
making it difficult to normalize the contribution of such spallation/fusion
reactions
to the observed/inferred Pop II lithium abundance.

Thus, the Pop II stellar data on lithium which is consistent with $[Li]_P =
2.2 \pm
0.2$ (Thorburn 1994), has strong assets but, a few liabilities.  On the
positive side,
 there is a
very large data base   ($\sim $ 100 stars) of accurate lithium abundances
in low metallicity
stars.  Looming on the negative side are the uncertain corrections for
reduction of the
surface lithium  in such old stars and for production of lithium via cosmic
ray nucleosynthesis.  The flatness of the Spite Plateau (Li vs. Fe) argues
against
these effects being large but, does not provide rigorous proof that their
contribution
is negligible.  It is, therefore, worthwhile to explore an alternate path
to primordial
lithium.  Observations of (or, searches for) interstellar (gas phase)
lithium in the
Galaxy and in the slightly less evolved LMC~ $([Fe/H]_{LMC} = - 0.3 \pm
0.1$; Russell
\& Bessell 1989) provide an indirect alternative.

\noindent INTERSTELLAR LITHIUM

There are many obstacles on the path to lithium abundances and lithium
evolution via
interstellar observations which account for this road being less traveled.
After
reviewing the major roadblocks I will propose a detour which leads to
constraints on the
abundance and evolution of lithium.

The first problem is observational.  The abundance of lithium is small $(
Li/H \sim
10^{-10} - 10 ^{-9})$ and the absorption features very weak; typical equivalent
widths
vary from a few tenths to a few  m\AA.  As a result, the number of
lines-of-sight with
absorption observed from ISM lithium is small.  With the current generation
of high
S/N, high resolution detectors, this obstacle can be overcome.

Even when lithium is observed in the ISM it is from LiI whereas lithium in
the ISM is
overwhelmingly LiII.  Thus, a large -- and uncertain -- ionization
correction must be
applied to derive Li II from the observed LiI.
$$ Li II/Li I = \Gamma_{Li}/\alpha _{Li} n _e. \eqno (1)$$
In (1), $\Gamma _{Li}$ is the LiI photoionization rate, $\alpha _{Li}$ is
the LiII
radiative recombination rate and $n _e$ is the (unobserved) local electron
density; the
shorthand notation LiII/LiI stands for the ratio of column densities.  To
determine
$n _e$, observations of CaII and CaI are often used.
$$ CaII/CaI = \Gamma _{Ca} / \alpha _{Ca} n _e. \eqno (2)$$
However, there are problems with this approach.  CaII is almost always
saturated
leading to a very uncertain determination of its column density, especially
of that part
of CaII which is coeval with the observed LiI.  Furthermore, we are
interested in
Li/H and the HI is also saturated, rendering the HI (which belongs to the
observed LiI)
uncertain.

Even when the large $(\sim 10^2 - 10^3)$ ionization correction is applied, the
inferred ISM abundance of lithium is found to lie well below (by 1-2 dex)
the solar
system value (Morton 1974; Snow 1975; Snell \& VandenBout 1981; White
1986).  It is
then usually    {\it assumed} that lithium is   {\it depleted} from the
gas phase of the ISM although the data alone do not distinguish between
depletion and a
true   {\it under}abundance (perhaps the Sun -- or, just the meterorites -- is
  {\it enhanced} in lithium (Steigman 1993)).  What follows, then, is a
tautology:  It is assumed that $(Li/H)_{ISM} = (Li/H)_{\odot}$ and, that the
difference between $ (Li/H)_{OBS} ~ {\rm and}~ (Li/H)_{\odot}$ is due to
depletion
onto grains and/or molecules
$$ [Li/H]_{OBS} \equiv -DF (Li)_{ISM} . \eqno (3)$$
In (3), and subsequently, the notation $[X/H]_A \equiv \log (X/H)_A - \log
(X/H)_{\odot}$ is used; thus, from (3), we have:  $ \log(Li/H)_{ISM} = \log
(Li/H)_{OBS} + DF (Li)_{ISM} = \log (Li/H)_{\odot}$.

Even if the DF(Li) could be calculated from first principles and, the
ionization
correction better constrained, the ISM derivations would be of limited value in
exploring the evolution of lithium.  The simple reason is that the ISM (of
the Galaxy)
is ``here and now".  To study the evolution of lithium requires that we
compare the ISM
abundance with that at an earlier epoch and/or lower metallicity.  Here
(finally!) the
road to primordial lithium improves.  SN87A   was bright enough to provide a
background source to probe the interstellar gas of the LMC.  Several groups
(Vidal-Madjar et al.\ 1987; Baade \& Magain 1988; Sahu, Sahu \& Pottasch
1988) searched,
unsuccessfully, for LiI absorption in front of SN87A.  Baade et al.\ (1991)
combined
all published data in hopes of extracting a signal but, were only able to
place an
upper bound on the LiI column density (at the LMC velocity).  This upper
bound to
N(LiI) contains important information on the evolution of lithium from the LMC~
$([Fe/H]_{LMC} \approx - 0.3 \pm 0.1$; Russell \& Bessell 1989) to the Galaxy
$([Fe/H]_{ISM} \approx 0)$.  But, as outlined above, great care must be
taken to avoid
the potholes of uncertain ionization and depletion corrections.

\noindent THE RELATIVE ABUNDANCE OF LITHIUM

Aside from any observational difficulties, the highly uncertain ionization and
depletion correction factors are a barrier to using ISM absorption data to
derive the
present abundance of lithium.  However, if the goal of deriving the
{\it absolute} abundance of lithium is deferred, the data can be utilized
to learn about its
  {\it relative} abundance.  Consider interstellar potassium.  As with
lithium, K in the ISM
is mainly KII  but, it is KI that is observed.  By comparing LiI to KI, a
much more accurate   {\it relative} abundance  Li/K can be obtained than the
separate   {\it absolute} (gas phase) abundances Li/H and K/H.
$$\biggl ({Li \over K}\biggr )_{OBS} = \biggl ({LiI \over KI}\biggr)_{OBS}
\biggl({\Gamma _{Li}/\Gamma _K
\over
\alpha _{Li}/ \alpha _K}\biggr ). \eqno (4) $$
The   {\it relative ionization correction factor},
$$icf(Li/K) \equiv \log \biggl ({\Gamma _{Li} / \Gamma _K \over \alpha
_{Li} / \alpha _K
}\biggr ),
\eqno (5)$$
is independent of the very uncertain electron density and, insensitive to
reasonable
variations in the photoionizing flux distribution.  Indeed, P\'equignot \&
Aldrovandi
(1986) consider four different radiation fields and find $ 5.9 \leq \Gamma
_{Li} /
\Gamma _K < 6.4$.  Allowing for this variation and, for ISM (HI region)
temperatures from
10K to 10$^3$K (P\'equignot \& Aldrovandi 1986),
$$ icf (Li/K) = 0.55 \pm 0.08. \eqno (6) $$
Especially significant in (6) is the small expected dispersion; aside from
an offset
(0.55 dex), $(Li/K)_{OBS}$ should correspond closely to $(LiI/KI)_{OBS}$.

However, until depletion is corrected for, $(Li/K)_{OBS}$ and
$(Li/K)_{ISM}$ need not
be the same.  Unless there is independent data, or a theory, to provide the
  {\it relative depletion}   {\it factor},
$$DF(Li/K) \equiv \log (Li/K)_{ISM} - \log (Li/K)_{OBS}, \eqno (7) $$
the ``true" ISM relative abundance remains unknown.

To see if this path is useful, consider the data (Hobbs 1984; White 1986).
There are
10 lines of sight (LOS) in the ISM where there are positive detections of
both LiI and
KI; the data (Hobbs 1984; White 1986) is presented in Table 1 and Figure 1.

Although the LiI and KI column densities each span an order of magnitude, the
  {\it ratio} has a dispersion of only 0.11 dex, which is smaller than the
typical ($1
\sigma)$ errors.  For the mean (either weighted or unweighted),
$$ \log (LiI/KI)_{OBS} = -2.43 \pm 0.04, \eqno (8) $$
where $\pm 0.04$ is the 1-sigma error in the mean.  Applying the relative
ionization
correction (eq. 6) to (8) yields the observed (i.e., gas phase ISM)
relative abundance,
$$ \log (Li/K)_{OBS} = -1.88 \pm 0.09, \eqno (9)$$
where the uncertainties in (6) and (8) have been combined in quadrature.

It is, perhaps, noteworthy that although relative column densities of NaI,
CaI and KI vary by
$ \sim 1 - 2$ dex along the   lines of sight   in Table 1, the LiI/KI ratio
shows no
statistically significant variation.  Even more interesting is the
comparison of $(Li/K)_{OBS}$
with $(Li/K)_{\odot}$ (Grevesse \& Anders 1989).
$$[Li/K]_{OBS} = \log (Li/K)_{OBS} - \log (Li/K)_{\odot} = - 0.06 \pm 0.10.
\eqno (10)$$
Thus, unless there is a cosmic conspiracy in which the observational data
and the inferred
ionization correction have arranged to cancel the relative depletion, the
result in (10)
strongly suggests that Li and K are similarly depleted in the ISM, as
anticipated in the models
of Snow (1975) and Field (1974).
$$DF(Li/K)_{ISM} = [Li/K]_{ISM} + 0.06 \pm 0.10. \eqno (11)$$
Since DF(Li/K) depends on the physics/chemistry of the gas phase depletion
while
\break
$(Li/K)_{ISM}$ depends on the stellar/galactic evolution of lithium and
potassium, it would
be surprising indeed that their difference is so small.  Although it must
be emphasized that this
can't be ``proven", it is not unreasonable to infer from (11) that
$DF(Li/K) \approx [Li/K]_{ISM}
\approx 0$ so that $(Li/K)_{ISM} \approx (Li/K)_{\odot}$.  Apparently, in
the last 4.6  Gyr, the
{\it relative} abundances of lithium and potassium have not changed much.

To recapitulate, the ratio of LiI and KI column densities along 10 LOS in
the ISM is observed
to be constant with a very small dispersion (0.1 dex), suggesting that, in
the local ISM,
neither the relative ionization correction nor the relative depletion
varies significantly from
place to place.  Thus, the observed ratio of column densities LiI/KI
provides a robust estimator
of the gas phase relative abundance Li/K.  When the relative ionization
correction (P\'equignot \&
Aldrovandi 1986) is applied, it is found that the gas phase abundance
$(Li/K)_{OBS}$ is very
close to   the solar system ratio, suggesting that $(Li/K)_{ISM} \approx
(Li/K)_{OBS} \approx (Li/K)_{\odot}$.  Next we turn to observations which
may provide a clue to
the evolution of this ratio.

\noindent  SN87A AND THE EVOLUTION OF LITHIUM

There is no primordial contribution to the abundance of potassium so that
its abundance at an
earlier epoch or in a less evolved system should be lower than that at
present.  Thus, for the
LMC it is expected that $(K/H)_{LMC} < (K/H)_{ISM}$ (ISM is used only for
the Galaxy).  Stellar
observations (Gratton \& Sneden 1987) provide support, suggesting that
potassium may scale
nearly linearly with metallicity $[K/H]
\sim  [Fe/H]$.  Lithium, in contrast, does have a BBN contribution and the
goal of this
analysis is to use the interstellar data to learn whether lithium has
started low (as suggested
by the observations of stars on the Spite Plateau for which $[Li/H]_{Pop
II} \approx -1.1 \pm
0.2)$ and is on its way up or, as required by much of parameter space for
inhomogeneous BBN, has
started high (as suggested by some models for depletion in rotating stars
(Pinsonneault,
Deliyannis \& Demarque 1991)) and is on its way down. If the primordial
lithium abundance were
comparable to or greater than the present (PopI) abundance, then for the
LMC it would be
expected that
$(Li/H)_{LMC}~ \rsim ~(Li/H)_{ISM}$ whereas for K which scales with Fe,
$(K/H)_{LMC} ~\lsim ~
(K/H)_{ISM}$. In this case we should expect to find that $(Li/K)_{LMC} >
(Li/K)_{ISM}$.
 The former case (Li on its way up) is more complicated, depending on the
relative lithium and potassium enhancements during the course of chemical
evolution.  To avoid
the biases of any specific model for chemical evolution and, to keep the
discussion as general
as possible, let us simply assume that the increase in lithium scales as a
power of the potassium
abundance:  $\Delta (Li/H)~ \sim (K/H)^{\alpha}$.  We may then write for $y
\equiv
(Li/H)/(Li/H)_{ISM}$ as a function of $x \equiv (K/H)/(K/H)_{ISM}$,
$$ y = A + (1 - A) x^{\alpha}, \eqno (12)$$
where $A \equiv (Li/H)_P/(Li/H)_{ISM}$.  For $\alpha ~\lsim ~1$ the Li to K
ratio (y/x)
{\it increases} with   {\it decreasing} potassium abundance (x).  In
contrast, if
$(y/x)_{LMC}$ should prove to be $< 1$, that would be a clear sign that
lithium has started low
and is rapidly on its way up.

The goal then is to compare $(Li/K)_{LMC}$ to $(Li/K)_{ISM}$.  To this end
it is   {\it \bf
not} necessary to compute Li/K with its attendant uncertain relative
ionization and depletion
corrections.  Rather, from the earlier discussions  it follows that
$$ \log (y/x)_{LMC} = \log
(LiI/KI)_{LMC} - \log (LiI/KI)_{ISM} +
\Delta icf + \Delta DF. \eqno (13)$$
In (13), $\Delta icf$ and $\Delta DF$ are, respectively, the   {\it
difference} in the
relative ionization correction and depletion factors from the LMC and the
ISM $(\Delta icf = icf
(Li/K)_{LMC} - icf (Li/K)_{ISM}$; $\Delta DF = DF (Li/K)_{LMC} - DF
(Li/K)_{ISM} $).  The
great virtue of (13) is that we may directly utilize the observational data
(LiI and KI
equivalent widths) and, we need not apply relative ionization correction
and depletion factors.
To infer y/x, or a bound to y/x, only requires that the sum of the   {\it
differences} in the
  {\it relative} icfs and DFs between the LMC and the ISM is small.  As
mentioned earlier,
P\'equignot and Aldrovandi (1986) find that the relative icf is insensitive
to the spectral
shape of the photoionizing flux so that $\Delta icf \approx 0$ is quite
reasonable.  Unlike
previous authors (e.g., Baade et al.\ 1991), we need not make any
assumptions about the absolute
depletion of lithium.  Rather, our only assumption is that the   {\it
relative} depletions
of Li and K in the LMC and in the ISM are similar $(\Delta DF \approx 0)$.
 From the study of dust
in the LMC (Fitzpatrick 1985), there is no evidence this is not a good
assumption.  Thus, we may
adopt,
$$ \log (y/x)_{LMC} \approx \log (LiI/KI)_{LMC} - \log (LiI/KI)_{ISM}.
\eqno (14)$$
In their comprehensive reanalysis of the searches for LiI absorption in the
LMC, Baade et al.\
(1991) also rederive the LMC KI equivalent width; their results correspond
to $ KI = 1.10 \times
10^{11} cm^{-2}$ or, $log (KI) = 11.04 \pm 0.02$.
Since no LiI absorption is detected at the LMC velocity, it is somewhat
difficult to assign a
statistical uncertainty (or confidence level) to the upper bound (Baade et
al. 1991). Although
the distribution of upper bounds to the LiI equivalent width is decidedly
non-gaussian, Baade
et al. (1991) find \break
\noindent $W_{\lambda} < 2.2 \pm 1.7 $m\AA \ which, at ``2$\sigma$" would
correspond to
$W_{\lambda} < 5.6 $m\AA. In fact, from Fig. 4 of Baade et al. (1991), 95\%
of the possible
$W_{\lambda}$ values have $W_{\lambda} < 5.6 $m\AA.  Thus, it is reasonable
to adopt a 95\% CL
upper bound of $W_{\lambda} < 5.6 $m\AA \ which corresponds to a ($\sim ``2
\sigma$") upper bound
on the column density of $\log (LiI)_{LMC} < 8.28$.  Since the uncertainty
in the upper bound to
the LMC LiI column density dominates that of the (observed) KI, we may
infer a $\sim$ 95\% CL
upper bound to the difference between LMC and ISM Li/K.
$$ \log (Li/K)_{LMC} < \log (Li/K)_{ISM} - 0.3 . \eqno (15)$$ This is the
key result of our
analysis.  The relative Li/K abundance in the LMC is
  {\it smaller,} by at least a factor of two, than the corresponding
relative abundance in
the ISM.  Lithium is on its way up from a primordial value  which is {\it
less} than its solar
system abundance.

\noindent TOWARDS THE PRIMORDIAL ABUNDANCE OF LITHIUM

The absence of LMC lithium absorption in the presence of LMC potassium
absorption (Baade et al.\
1991), when combined with the Galactic LiI and KI data, argues for a
primordial abundance of
lithium less -- by at least a factor of $ \sim 2$ -- than the present Pop I
lithium abundance.
How much less depends on assumptions which range from eminently reasonable
to speculative.  Let
us begin with the most reasonable -- and, therefore, least constraining --
assumption.

Although potassium is not observed in LMC stars, it is reasonable to assume
that $(K/H)_{LMC}
< (K/H)_{GAL} ~({\rm i.e.,}\ x_{LMC} < 1)$.  Therefore, $[Li]_P \leq
[Li]_{LMC} < [Li]_{GAL} -
0.3$. The solar system lithium abundance $( [Li]_{\odot} = 3.31 \pm
0.04)$,  and the stellar abundances $([Li]_{PopI} = 3.2 \pm 0.1$) derived
from T-Tauri stars
(suiitably corrected for veiling and NLTE; Magazz\'u, Rebolo \& Pavlenko
1992) and
hot stars in young open clusters (Boesgaard \& Tripico 1986; Balachandran
1995) suggest a $2
\sigma$ upper bound of
$[Li]_{GAL} \le 3.4$.  Thus at $> 95\%$ CL we may infer   a ``zeroth order"
upper bound to
primordial lithium,
$$[Li]{^{(0)} _P} < 3.1 .\eqno (16)$$
This zeroth order bound is conservative in the sense that potassium
decreases with
metallicity (Gratton \& Sneden 1987) and $[Fe/H]_{LMC} \sim -0.3$  (Russell
\& Bessell 1989).
Thus, for $[K/H]_{LMC} \sim -0.3$, we may infer a more reasonable ``first
order"
bound to primordial lithium,
$$[Li]{_P ^{(1)}} < 2.8. \eqno (17)$$

Finally, we may use the ``generic" variation of Li vs. K   described by eq. 12
to bound $[Li]_P$ (where: $[Li]_P = \log A + [Li]_{GAL} < 3.4 + \log A$).
Since $(y/x)_{LMC}\
\lsim 1/2$, it follows that K must decrease with metallicity less rapidly
than Li; i.e., $\alpha
> 1$ (where $\Delta (Li/H) \sim (K/H)^{\alpha})$.  Indeed, for $[K/H]_{LMC}
>\ \sim [Fe/H]_{LMC}
\ \rsim - 0.5, \alpha >1.6$.  An obstacle to employing eq. 12, along with
$(y/x)_{LMC}\ \lsim
1/2$, to infer a bound to A is that $x_{LMC}$ is  not observed directly.
However,
$x_{LMC}$ is, in fact, not needed to derive an upper bound to A since, for
$\alpha >1$,
$$ A = \Big [(y/x)_{LMC}\ x_{LMC}  - x {^{\alpha} _{LMC}}\Big ]/(1 - x{
^{\alpha} _{LMC}}), \eqno
(18)$$    is maximized for   {\it some} value of $x_{LMC} < 1$.  For
example, for $\alpha =
2$, $A_{MAX} = \break 1/2 [ 1 - (1 -(y/x){^2 _{LMC}})^{1/2}]$ which, for $
(y/x)_{LMC} < 1/2,
 A_{MAX} < 0.067$ (or, $\log A_{MAX} < - 1.2$).    This leads to a very
strong upper bound to
primordial lithium,
$$[Li]{_P ^{(2)}} < 2.2 . \eqno (19)$$
Since $A_{MAX}$ increases with increasing $\alpha$, we should perhaps, regard
this result (19)
with some caution.  In general, $$ A_{MAX}\ \lsim {\Big ( {\alpha - 1 \over
\alpha} \Big ) \Big (
{y \over x} \Big )}^{(\alpha / \alpha - 1)} _{LMC} , \eqno (20)$$
so that for $\alpha = 3$, $A_{MAX}\ \lsim 0.24$ and
$$[Li]{^{(3)} _{P}}\ \lsim 2.8. \eqno (21)$$
Thus, although the relatively low upper bound to $(Li/K)_{LMC}$ suggests a
quite small abundance
of primordial lithuim (eq. 19) which, by the way, is entirely consistent
with the stellar data
(Molaro et al. 1995), the bounds in eqs. 17 and 21 provide a relatively
firm upper bound of
$[Li]_P \ \lsim 2.8$.  This bound is entirely consistent with the Pop II
data (Thorburn 1994;
Molaro et al.
1995), even allowing for stellar depletion/dilution (Chaboyer et al. 1992;
Pinsoinneault, Deliyannis \& Demarque 1992).

\noindent SUMMARY

Observations of lithium on the surface of young (Pop I) and old (Pop II)
stars provide a
valuable probe of the abundance of lithium and its evolution.  However, in
$\sim 10 - 15$ Gyr of
evolution, the oldest stars may have modified their surface abundances and,
therefore, no longer
offer a reliable probe of the lithium abundance in the nearly primordial
gas out of which they
formed.  Indeed, evidence for large depletion exists for relatively cool
Pop I and Pop II
stars.  However, the flatness of the ``Spite Plateau" argues that the
lithium abundance, derived
from observations of the warm $( \rsim 5700 K)$, metal-poor $([Fe/H] ~\lsim
-1.3)$ stars,
represents the primordial value which has been little modified by stellar
depletion or early
Galaxy production mechanisms.  Although reasonable, this common assumption
cannot be proven
rigorously.  Indeed, the rotating stellar models of Pinsonneault,
Deliyannis \& Demarque (1992)
suggest that large reductions in prestellar lithium are not inconsistent
with the data.

As a complement to the stellar observations, it has been suggested here
that interstellar data can
also provide valuable information on the evolution of lithium and, may be
used to bound the
primordial abundance from above.  Extreme caution must be exercised in
utilizing interstellar
observations if practical and logical pitfalls are to be avoided.  Large
and possibly uncertain
ionization corrections (to infer LiII from the observed LiI) must be
applied, as well as
unknown (even unknowable!) depletion factors (to go from the gas phase to
total interstellar
abundances).  To avoid these traps, it was proposed here that the goal of
the absolute
abundance (Li/H) be replaced with the more modest target of the relative --
to potassium --
abundance (Li/K).  Even so, to proceed from the observed LiI/KI to the
derived Li/K still
requires knowledge of the  {\it relative} ionization correction factor
(icf(Li/K)) and
the   {\it relative} depletion factor (DF(Li/K)).  Here, this approach has
been first
applied to data from 10 lines-of-sight in the Galaxy.  The small dispersion
($\sim 0.1 dex)$
around a mean LiI/KI ratio suggests that the relative icf and DF do not
vary much -- if at all
-- in the local ISM.  And, indeed, when the theoretical relative icf
(P\'equignot \& Aldrovandi
1986) is used, the data lead to $(Li/K)_{ISM} \approx (Li/K)_{\odot}$,
suggesting that the
relative depletion factor $DF(Li/K)_{ISM} \approx 0$.

\vfil\eject
This complementary approach of relative abundances is especially valuable
for probing the
evolution of lithium.  SN87A provided a bright source to use -- only
briefly! -- in studying the
interstellar gas of the LMC.  When comparing $(Li/K)_{LMC} ~{\rm to}~
(Li/K)_{GAL}$, it is only
the   {\it differences} in   {\it relative} icfs and DFs that enter.  On
the reasonable
assumption that $ \Delta icf (Li/K) + \Delta DF (Li/K) \approx 0$, the
difference between Li/K
in the slightly less evolved LMC and in the Galaxy follows directly from
the Li I and K I
observations.
$${(Li/K)_{LMC} \over (Li/K)_{GAL}} \approx { (LiI/KI)_{LMC} \over
(LiI/KI)_{GAL}} < 1/2. \eqno
(22)$$ That this ratio is less than unity establishes that lithium is
evolving from a primordial
abundance less (by at least a factor of 2) than its present value.  This
qualitative constraint
is nonetheless sufficient to rule out much of parameter space for models of
inhomogeneous BBN
(Thomas et al.\ 1994) and to limit the depletion/dilution of lithium in Pop
II stars (Chaboyer et
al.\ 1992; Pinsonneault, Deliyannis \& Demarque 1992).

Since the abundance of lithium has a ``floor" -- its primordial (BBN) value
-- while potassium
does not, the upper bound to Li/K in the less evolved LMC, permits us to
derive an upper bound
to primordial lithium.  If it is only assumed that $ (K/H)_{LMC} \leq
(K/H)_{GAL}$, we have found
$[Li]{_P ^{(0)}} < 3.1$ As a next approximation we have used $[Fe/H]_{LMC}$
to infer
$[K/H]_{LMC}$
  leading to $[Li]{_P ^{(1)}} < 2.8$.  Finally, by a ``generic" scaling of
the lithium production to that of potassium $(\Delta (Li/H) \sim
(K/H)^{\alpha}, ~ \alpha \geq
2)$, we have used the LMC bound to provide an even tighter constraint on
primordial lithium:
$[Li]{_P ^{(2)}} ~ < 2.2\ ( {\rm or,}~ [Li]{_P ^{(3)}} < 2.8$).

The relatively low upper bounds to $[Li]_P$ derived here from interstellar
absorption data lend
support to the notion that the ``Spite Plateau" abundance provides a fair
estimate of the
primordial abundance.  There is not much room for significant modification
of the observed Pop
II abundances by stellar surface depletion and/or cosmic ray
nucleosynthesis.  For example,
Thorburn's (1994) estimate $[Li]_P = 2.22 \pm 0.20 ~(2 \sigma)$ is entirely
consistent with all
the upper bounds presented above.  Such a low value for primordial lithium
leads to a
significant upper bound to the universal abundance of nucleons ($\eta_{10}
= 10^{10}~(N/\gamma))$
and, hence, to the present density contributed by them $(\Omega _B = 0.015
\eta _{10} h{^{-2}
_{50}}, ~ H_0 = 50 h_{50} kms^{-1}Mpc^{-1})$.  For example, for $[Li]_P\
\lsim 2.8, \eta _{10}\
\lsim  8 \ {\rm and \ } \Omega _B h{^2 _{50}}\ \lsim  0.12$.

Stellar and interstellar observations provide complementary probes of
lithium and its
evolution.  Although subject to vastly different observational and physics
uncertainties, they
yield a consistent picture of primordial lithium supporting the conclusion
that the warm,
metal-poor Pop II stars (Spite Plateau) provide a fair estimate of the BBN
abundance.  It is now
quite feasible, and it would be very valuable, to increase the number of
lines-of-sight in the
Galaxy with Li I and K I observations.  Even more interesting would be to
use future, bright
supernovae in other galaxies to probe the evolution of lithium.

\noindent
ACKNOWLEDGEMENTS

During the long gestation period of this work I have profited from
informative discussion with
and encouragement from many colleagues.  I wish to thank especially D.
Baade, L. Hobbs, B.
Pagel, M. Pinsonneault and S. Viegas.  Much of this work was done while the
author was an
overseas Fellow at Churchill College and a Visiting Scientist at the
Institute of Astronomy
(Cambridge, England) and he thanks them for hospitality.  This work is
supported at OSU by DOE
grant DE-FG02-94ER-40823.

\vfil\eject

\begin{center}
\begin{tabular}{ c c c c }
\multicolumn{4}{c} {TABLE 1}\\
\multicolumn{4}{c}{Galactic Li I and K I$^{\S}$}\\
\\
\hline
\hline\\
\underbar {Line of Sight} & \underbar {$log (LiI)^{\ast}$} & \underbar{log
(KI)}
&\underbar{log (LiI/KI)} \\
\\
\hline
\\


$\delta $ Sco & \ \ 8.92 $\pm$ 0.11$^{\star}$ & 11.40 $\pm$ 0.20 & -2.48
$\pm$ 0.23 \\
$\sigma $ Sco & \ 9.32 $\pm$ 0.15 & 11.53 $\pm$ 0.25 & -2.21 $\pm$ 0.29\\
$ \zeta$ Oph &\ 9.37 $\pm$ 0.09 & 11.87 $\pm$ 0.12 & -2.50 $\pm$ 0.15\\
$\epsilon$ Aur &\ 9.41 $\pm$ 0.10 & 11.88 $\pm$ 0.20 & -2.47 $\pm$ 0.22 \\
$\zeta $ Per &\ 9.46 $\pm$ 0.07 & 11.90 $\pm$ 0.17 & -2.44 $\pm$ 0.18   \\
${\chi}^2$ Ori &\ 9.70 $\pm$ 0.27 &12.18 $\pm$ 0.13 & -2.48 $\pm$ 0.30 \\
55 Cyg &\ 9.72 $\pm$ 0.12 & 12.04 $\pm$ 0.20 & -2.32 $\pm$ 0.23 \\
$\rho$ Oph &\ 9.94 $\pm$ 0.07 & 12.23 $\pm$ 0.28 & -2.29 $\pm$ 0.29 \\
$\eta$ Cep &\ 9.94 $\pm$ 0.25 & 12.52 $\pm$ 0.33 & -2.58 $\pm$ 0.41\\
HR7573 &10.00 $\pm$ 0.21 & 12.50 $\pm$ 0.34& -2.50 $\pm$ 0.40 \\
\hline

\end{tabular}
\end{center}
 $^{\S}$ Data from Hobbs (1984) and White (1986).

\noindent $^{\ast}$ The Li I and K I column densities in cm$^{-2}$.

 \noindent $^{\star}$ The 2-sigma statistical uncertainties; the errors
have been
combined in quadrature \break \hspace*{.19in}for LiI/KI.

 \noindent REFERENCES

   \noindent Alcock, C. R., Fuller, G \& Mathews, G. J. 1987, {\it ApJ},
{\bf 320}, 439\\
    Applegate, J. H., Hogan, C. \& Scherrer, R. J. 1988, {\it ApJ}, {\bf
329}, 572\\
    Baade, D. \& Magain, P. 1988, {\it A \& A}, {\bf 194}, 237\\
   Baade, D., Cristiani, S., Lanz, T., Malaney, R.A., Sahu, K. C. \&
Vladilo, G. 1991, {\it
A \& A}, \indent{\bf 251}, 253\\
  Balachandran, S. 1995,   {\it Ap. J.} In Press (June 1995)  \\
   Boesgaard, A. M. \& Steigman, G. 1985, {\it A \& A} {\bf 23}, 319\\
    Boesgaard, A. M. \& Tripico, M. J. 1986, {\it ApJ}, {\bf 303}, 724\\
 Chaboyer, B., Deliyannis, C. P., Demarque, P., Pinsonneault, M. H. \&
Sarajedini, A.
1992, \indent{\it ApJ}, {\bf 388}, 372\\
   Charbonnell, C. \& Vauclair, S. 1992, {\it A \& A}, {\bf 265}, 55\\
  Duncan, D. K., Lambert, D. L. \& Lemke, M. 1992, {\it ApJ}, {\bf 401}, 584\\
  Field, G. B. 1974, {\it ApJ}, {\bf 187}, 453\\
  Fitzpatrick, E. L. 1985, {\it ApJ}, {\bf 299}, 219\\
  Gilmore, G., Edvardsson, B. \& Nissen, P. E. 1992, {\it AJ}, {\bf 378}, 17\\
  Gratton, R. G. \& Sneden, C. 1987, {\it A \& A}, {\bf 178}, 179\\
  Grevesse, N. \& Anders, E. 1989, in {\it ATP Conf. Proc. 183, Cosmic
Abundances of
Matter}, \indent ed. C. J. Waddington (New York; AIP), 1\\
  Hobbs, L. M. 1984, {\it ApJ}, {\bf 286}, 252\\
  Hobbs, L. M. \& Pilachowski, C. 1988, {\it ApJ}, {\bf 326}, L23\\
  Hobbs, L. M. \& Thorburn, J. A. 1994, {\it ApJ}, {\bf 428}, L25\\
  Kurki-Suonio, H. \& Matzner, R. 1990, {\it Phys. Rev. D}, {\bf 42}, 1046\\
  Magazzu, A. Rebolo, R. \& Pavlenko, Y. V. 1992, {\it ApJ}, {\bf 392}, 159\\
  Malaney, R. A. \& Fowler, W. A. 1988, {\it ApJ}, {\bf 33}, 14\\
  Mathews, G. J., Meyer, B. S., Alcock, C. R. \& Fuller, G. M. 1990, {\it
ApJ}, {\bf 358}, 36\\
Molaro, P., Primas, F. \& Bonifacio, P. 1995, {\it Astron. Astrophys.
Lett}. (In Press)\\
Morton, D. C. 1974 {\it ApJ}, {\bf 193}, L35\\
  P\'equignot, D. \& Aldrovandi, S. M. V. 1986, {\it A \& A}, {\bf 161}, 169\\
  Pinsonneault, M. H., Deliyannis, C. P. \& Demarque, P. 1992, {\it ApJS},
{\bf 78}, 179\\
  Rebolo, R., Molaro, P. \& Beckman, J. 1988, {\it A \& A}, {\bf 192}, 192\\
  Russell, S. C. \& Bessell, M. S. 1989, {\it ApJS}, {\bf 70}, 865\\
  Ryan, S., Bessel, M., Sutherland, R. \& Norris, J. 1990, {\it ApJ}, {\bf
348}, L57\\
  Ryan, S., Norris, J., Bessel, M. \& Deliyannis, C. 1992, {\it ApJ}, {\bf
388}, 184\\
  Sahu, K. C., Sahu, M. \& Pottasch, S. R. 1988, {\it A \& A}, {\bf 207}, L1\\
  Smith, M. S., Kawano, L. H. \& Malaney, R. A. 1993, {\it ApJS}, {\bf 85},
219\\
  Smith, V. V., Lambert, D. L. \& Nissen, P. E. 1992, {\it ApJ}, {\bf 408},
262\\
 Snell, R. L. \& VandenBout, P. A. 1981, {\it ApJ}, {\bf 250}, 160\\
  Snow, T. P. 1975, {\it ApJ}, {\bf 202}, L87\\
  Spite, F., Maillard, J. P. \& Spite, M. 1984, {\it A \& A}, {\bf 141}, 56\\
 Spite, M. \& Spite, F. 1982, {\it A \& A}, {\bf 115}, 357; {\it Nature},
{\bf 297}, 483\\
  Spite, F. \& Spite, M. 1986, {\it A \& A}, {\bf 163}, 140\\
  Steigman, G. 1993, {\it ApJ}, {\bf 307}, 777\\
  Steigman, G. \& Walker, T. P. 1992, {\it ApJ}, {\bf 385}, L13\\
  Steigman, G., Fields, B. D., Olive, K. A., Schramm, D. N. \& Walker, T.
P. 1993, {\it
ApJ}, {\bf 415}, \indent L35\\
  Terasawa, N. \& Sato, K. 1989, {\it Progr. Theor. Phys. Letters}, {\bf
81}, 254\\
 \newpage \noindent Thomas, D., Schramm, D. N., Olive, K. A. Mathews, G.,
Meyer, B. S., \& Fields
B. D. 1994,
\indent {\it ApJ}, {\bf 430}, 391\\
  Thorburn, J. A. 1994, {\it ApJ}, {\bf 421}, 318\\
   Vidal-Madjar, A., Andreani, P., Cristiani, S., Ferlet, R., Lanz, T. \&
Vladilo, G. 1987,
\break \indent{\it A \& A},   {\bf 177}, L17\\
  Walker, T. P., Steigman, G., Schramm, D. N., Olive, K. A. \& Kang, H.-S,
1991, {\it ApJ},
{\bf 376}, \indent 51\\
White, R. E. 1986, {\it ApJ}, {\bf 307}, 777\\

\vfil\eject

\noindent Figure Caption

\noindent Figure 1.  The log of the LiI/KI column density ratios for 10
lines-of-sight in the
Galaxy (from Hobbs 1984 and White 1986); the error bars are $\pm 2 \sigma$.
Also shown is the
solar system (meteoritic) value (Grevesse \& Anders 1989) ``corrected" by
the icf $= 0.55 \pm
0.08$ and, the
$2
\sigma$ upper bound from the LMC (Baade et al.\ 1991).

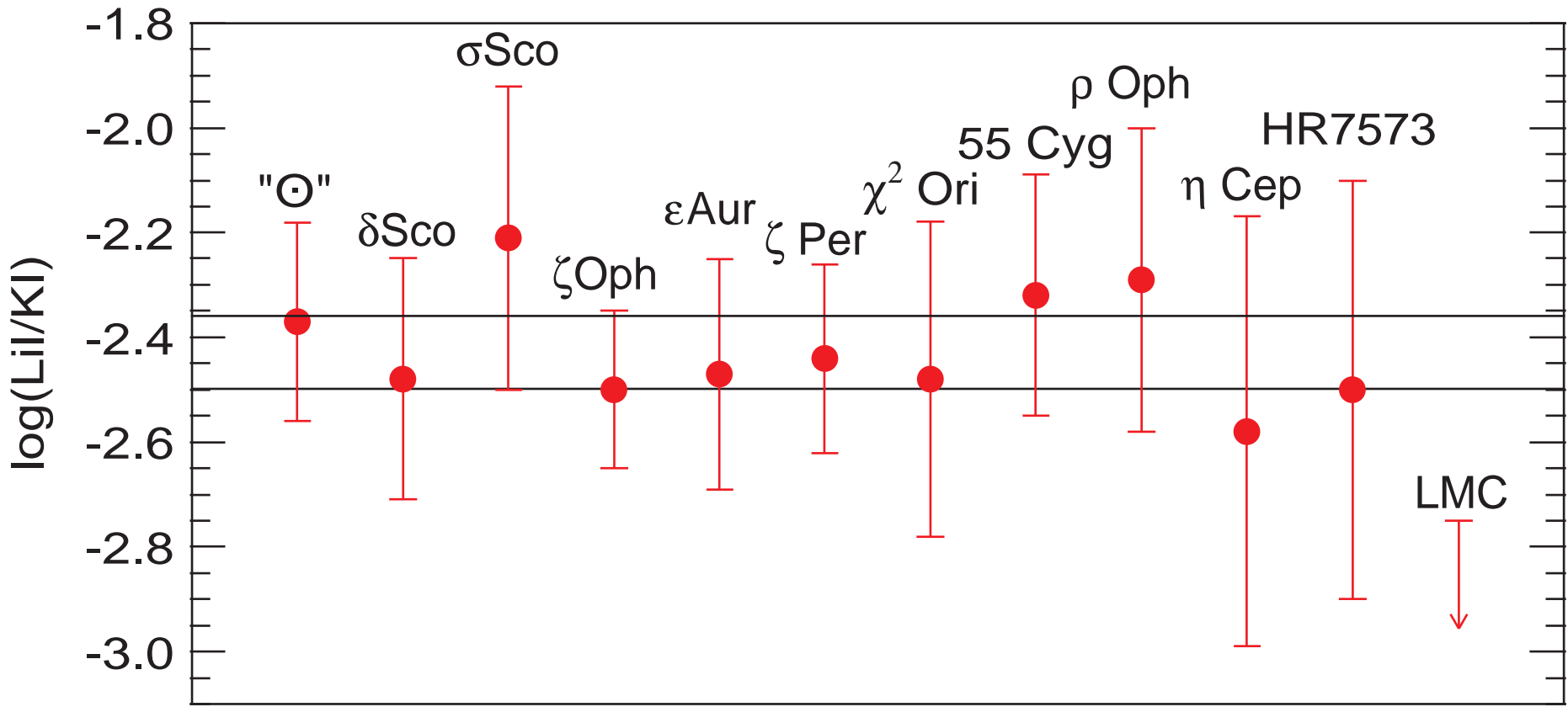
\end{document}